\documentstyle[aps,preprint,prb]{revtex}
\begin{document}
\title{\bf Pairing Constraint on the Real Space Formalism of \\
the Theory of Superconductivity$^{\dagger}$}
\author{Yong-Jihn Kim$^{\dagger\dagger}$}
\address{Department of Physics, Purdue University, West Lafayette, Indiana 47907}
\maketitle
\begin{abstract}
We review the new theory of impure superconductors 
constructed by Kim and Overhauser, and further developed by Kim. 
It was shown that Gor'kov's self-consistency equation needs a pairing
constraint derived from the Anomalous Green's function. 
Whereas earlier studies have applied a constraint only on the
pair potential, we show that the kernel also should be fixed by the 
ground state.
The Bogoliubov-de Gennes equations need a pairing constraint in 
order to obtain the proper vacuum state by the corresponding unitary 
transformation. 
The relation between the pair potential and the gap parameter is clarified.
This new study opens up a reinvestigation of inhomogeneous superconductors.
We will discuss (i) strong coupling theory of dirty superconductors,
(ii) suppression of magnetic impurity effect by ordinary
impurities, (iii) weak localization correction to the
phonon-mediated interaction, and (iv) other inhomogeneous
superconductors.
\end{abstract}
\vskip 1pc
\noindent
$^{\dagger}$ Invited Talk at the Inauguration Conference of Asia Pacific
Center for 

Theoretical Physics, June 4-10, 1996, Seoul, Korea

\noindent
$^{\dagger\dagger}$ Present address: Department of Physics, Korea Advanced 
Institute of Science 

and Technology, Taejon 305-701, Korea
\vfill\eject
\section{Introduction}
Recently it was shown$^{1,2}$ that  
the Abrikosov-Gor'kov (AG)$^{3}$ theory of impure superconductors
predicts a large decrease of $T_{c}$, linear in the 
nonmagnetic impurity concentration, which is
not consistent with Anderson's theorem.$^{4}$ 
In their response$^{5}$ on ref. 1, AG argued that the frequency cutoff makes the AG theory 
compatible with Anderson's theorem. 
This argument is based on Tsuneto's application$^{6}$ of the AG theory
to the Eliashberg equation.

To settle this controversy, we first need to understand 
the limitation of Anderson's theorem.
It was pointed out that Anderson's theorem is 
valid only up to the first order of the impurity concentration,$^{1,7}$ 
and the phonon-mediated interaction is strongly decreased by Anderson
localization.$^{1,8,9}$

However, Tsuneto's theory fails to show the existence of localization
correction to the phonon-mediated interaction.
The failure comes from the intrinsic pairing problem$^{10,11}$ in Gor'kov's
formalism.$^{12}$ 
The kernel of the self-consistency equation should be set by the
physical constraint of the Anomalous Green's function. 
The resulting equation is nothing but another form
of the BCS gap equation.
Using the equation, the localization correction
to the phonon-mediated interaction may be calculated.$^{1,11}$
A correct strong coupling theory has been reported by Kim.$^{8}$

For magnetic impurity effects, Kim and Overhauser (KO)$^{7}$ proposed a BCS type
theory with different predictions: (i) The initial slope of $T_{c}$
decrease depends on the superconductor and is not the universal
constant proposed by Abrikosov and Gor'kov(AG).$^{3}$ (ii) The $T_{c}$ reduction
by exchange scattering is partially suppressed by potential scattering
when the overall mean free path is smaller than the coherence length.
This compensation has been confirmed in several experiments.$^{13-15}$
The difference comes again from the pairing problem.
If we impose a correct pairing condition
on the self-consistency equation, or the AG's calculation, we can find KO's result.

\section{Phonon-Mediated Interaction in BCS Theory and Gor'kov's Formalism }
\subsection{BCS Theory}

For a homogeneous system, BCS introduced 
a reduced version of the Fr\"ohlich phonon-mediated interaction,
\begin{eqnarray}
H_{red} = \sum_{{\vec k} {\vec k}'} V_{{\vec k} {\vec k}'} c_{{\vec k}'}^{\dagger}c_{-{\vec k}'}^{\dagger}
c_{-{\vec k}}c_{{\vec k}}, 
\end{eqnarray}
where
\begin{eqnarray}
V_{{\vec k}{\vec k}'}= \cases{-V, &if $|\epsilon_{{\vec k}}|,|\epsilon_{{\vec k}'}|\leq  \omega_{D}$\cr
                      0, &otherwise.\cr}
\end{eqnarray}
This reduction procedure is recognizing in advance which eigenstates
will be paired and so contribute to the BCS condensate.

In the presence of impurities, we can derive the phonon-mediated 
interaction by transforming the Fr\"ohlich interaction 
using the relation, 
\begin{eqnarray}
\psi_{n\sigma} = \sum_{{\vec k}}\phi_{{\vec k}\sigma}<{\vec k}|n>.
\end{eqnarray}
$\psi_{n}$ and $\phi_{{\vec k}}$ denote the scattered state and
the plane wave state.
The reduced version of this interaction anticipates that $\psi_{n}$ (having
spin up) will be paired with its time-reversed counterpart 
$\psi_{\overline {n}}$ (having spin down). The new reduced Hamiltonian is
\begin{eqnarray}
H_{red}' = \sum_{n n'} V_{n n'}c_{n'}^{\dagger}c_{\overline {n}'}^{\dagger}c_{\overline
{n}}c_{n},  
\end{eqnarray}
where 
\begin{eqnarray}
V_{n n'} = -V\sum_{{\vec k} {\vec k}' {\vec q}}  <{\vec k} - {\vec q}|n'>
<{\vec k}' + {\vec q}|\overline {n}'><{\vec k}'|\overline {n}>^{*}<{\vec k}|n>^{*}. 
\end{eqnarray}
Anderson's theorem is valid only when  ${\vec k}'$ can be set equal to 
$-{\vec k}$. 

\subsection{Gor'kov's formalism}

In Gor'kov's formalism, a point
interaction $-V\delta({\bf r}_{1}-{\bf r}_{2})$ is used for the 
pairing interaction between electrons. 
For a homogeneous system, the pairing interaction is
\begin{eqnarray}
H_{G} &=& - {1\over 2}V\int d{\bf r}\sum_{\alpha\beta}\Psi^{\dagger}({\bf r}\alpha)
\Psi^{\dagger}({\bf r}\beta)\Psi({\bf r}\beta)\Psi({\bf r}\alpha) \nonumber \\
&=&-{1\over 2}V\sum_{{\vec k}{\vec k}'{\vec q}\sigma\sigma '}
c_{{\vec k} - {\vec q}, \sigma}^{\dagger}c_{{\vec k}' + {\vec q}, \sigma '}^{\dagger}c_{{\vec k} '\sigma '}
c_{{\vec k}, \sigma}, 
\end{eqnarray}
and
\begin{eqnarray}
V_{{\vec k}{\vec k}'}&=&-V\int \phi_{{\vec k}'}^{*}({\bf r}) \phi_{-{\vec k}'}^{*}({\bf r})
 \phi_{-{\vec k}}({\bf r}) \phi_{{\vec k}}({\bf r})d{\bf r}\nonumber \\
&=&-V.
\end{eqnarray}
Eq. (6) is the same as the Fr\"ohlich interaction within the BCS
approximation.

Note that the two points are not clear in Gor'kov's formalism, i.e.,
the BCS reduction procedure and the retardation cutoff.
To obtain the same result as that of the BCS theory,
these two ingredients should be taken care of in some way.
As will be shown later, the negligence of the BCS reduction procedure
causes a serious pairing problem especially in impure superconductors. 

In the presence of impurities, 
the matrix element of the pairing interaction is
\begin{eqnarray}
V_{nn'}=-V\int \psi_{n'}^{*}({\bf r}) \psi_{\bar n'}^{*}({\bf r})
 \psi_{\bar n}({\bf r}) \psi_{n}({\bf r})d{\bf r}.
\end{eqnarray}
Substituting Eq. (3) into Eq. (8) we find that
\begin{eqnarray}
V_{nn'}=-V \sum_{{\vec k} {\vec k}' {\vec q}} <{\vec k} - {\vec q}|n'>
<{\vec k}' + {\vec q}|\overline {n}'><{\vec k}'|\overline {n}>^{*}<{\vec k}|n>^{*}. 
\end{eqnarray}
Notice  that Eq. (9) is the same as Eq. (5).
\section{Pairing Constraint on Gor'kov's Formalism}
\subsection{Inhomogeneous System: Nonmagnetic Impurity Case}
Near the transition temperature, 
the usual self-consistency equation is
\begin{eqnarray}
\Delta({\bf r}) &=& VT\sum_{\omega}\int \Delta({\bf l})G^{\uparrow}_{\omega}({\bf r,l})
G^{\downarrow}_{-\omega}({\bf r,l})d{\bf l}\nonumber \\
&=&\int K({\bf r},{{\bf l}})\Delta({{\bf l}})d{{\bf l}}.
\end{eqnarray}
Note that Kernel $K({\bf r},{{\bf l}})$ is not for Anderson's
pairing.
It includes the extra pairings between $n\uparrow$ and $n'(\not={\bar n})\downarrow$. 
The kernel for Anderson's pairing is 
\begin{eqnarray}
K^{A}({\bf r},{{\bf l}})&=&VT\sum_{\omega}
\{G^{\uparrow}_{\omega}({\bf r}, {{\bf l}}) 
  G^{\downarrow}_{-\omega}({\bf r'}, {{\bf l}})\}_{p.p.}\nonumber \\
&=&V\sum_{n}{1\over 2\epsilon_{n}}tanh{\epsilon_{n}\over 2T}\psi_{n}({\bf r})
\psi_{\bar n}({\bf r}) \psi^{*}_{\bar n}({{\bf l}})\psi^{*}_{n}({{\bf l}}),
\end{eqnarray}
where p.p. means proper pairing constraint, which dictates
pairing between $n\uparrow$ and ${\bar n}\downarrow$.

It can be shown$^{10,11}$ that the extra pairings violate the physical 
constraint of the Anomalous Green's function, i.e.,
\begin{eqnarray}
\overline{F({\bf r},{\bf r'},\omega)}^{imp}&\sim& 
\overline{\psi_{n\uparrow}({\bf r})\psi_{n'\downarrow}({\bf r'})}^{imp}\nonumber \\
&\not=& \overline{F({\bf r}-{\bf r'},\omega)}^{imp}.
\end{eqnarray}
These extra pairings should have been eliminated by the BCS reduction
procedure in the Hamiltonian. 
Consequently,  the revised  self-consistency equation is
\begin{eqnarray}
\Delta({\bf r}) = VT\sum_{\omega}\int \Delta({{\bf l}})\{G^{\uparrow}_{\omega}({\bf r,{\bf l}})
G^{\downarrow}_{-\omega}({\bf r,{\bf l}})\}_{p.p.}d{\bf l}.
\end{eqnarray}
Notice that Eq. (13) is nothing but another form of the
BCS gap equation,
\begin{eqnarray}
\Delta_{n}=\sum_{n'}V_{nn'}{\Delta_{n'}\over 2E_{n'}}tanh
{E_{n'}\over 2T}.
\end{eqnarray}

\subsection{Inhomogeneous System: Magnetic Impurity Case}

KO's $^{7}$ theory employed degenerate
scattered state pairs.   
It has been claimed that the inclusion of the extra pairing 
is the origin of the 
so-called pair-breaking of the magnetic impurities.$^{16,17}$
However, the extra pairing terms cause the violation of the physical 
constraint of the pair potential and  the Anomalous Green's function.$^{11}$
It can be shown$^{11}$ that the homogeneity condition of the
Anomalous Green's function, after the impurity average, requires pairing 
between the
degenerate scattered state partners. Then the revised self-consistency equation
gives rise to  KO's result.$^{7}$

\subsection{Homogeneous System}

Near the transition temperature, the Anomalous Green's function is given by
\begin{eqnarray}
F({\bf r}, {\bf r'}, \omega) = \int \Delta({{\bf l}})G^{\uparrow}_{\omega}({\bf r}, {{\bf l}})
G^{\downarrow}_{-\omega}({\bf r'}, {{\bf l}})d{{\bf l}},
\end{eqnarray}
Gor'kov$^{12}$ pointed out that $F({\bf r},{\bf r'})$ should depend only on ${\bf r}-{\bf r'}$, i.e.,
\begin{eqnarray}
F({\bf r},{\bf r'}) = F({\bf r}-{\bf r'}).
\end{eqnarray}
Note that Eq. (15) includes the extra pairing terms
between ${\vec k}\uparrow$ and ${\vec k}'\downarrow(\not= -{\vec k}\downarrow)$, which do not
satisfy the homogeneity condition of Eq. (16). In this case, the self-consistency
condition of the pair potential happens to eliminate the extra pairing
in Eq. (15) because of the orthogonality of the wavefunctions.
 
However, it is important to eliminate the extra pairing
in the Anomalous Green's function from the beginning.
Note that the kernel $K({\bf r},{{\bf l}})$ is not for the pairing 
between ${\vec k}\uparrow$ and $ -{\vec k}\downarrow$, but for the pairing between
the states which are the linear combination of the plane
wave states $\phi_{{\vec k}}({\bf r})$.$^{18}$ 
The inclusion of the extra pairings
hindered our correct understanding of the 
impure superconductors and the relation between the
pair potential and the gap parameter.  
\section{Pairing Constraint on the Bogoliubov-de Gennes Equations}
\subsection{Inhomogeneous system: Nonmagnetic Impurity Case}
By performing a unitary transformation, 
\begin{eqnarray}
\Psi({\bf r}\uparrow) & =& \sum_{n}(\gamma_{n\uparrow}u_{n}({\bf r}) - 
\gamma^{\dagger}_{n\downarrow}v^{*}_{n}({\bf r})) \nonumber \\
\Psi({\bf r}\downarrow) & =& \sum_{n}(\gamma_{n\downarrow}u_{n}({\bf r}) + 
\gamma^{\dagger}_{n\uparrow}v^{*}_{n}({\bf r})),
\end{eqnarray}
 we obtain the well-known  Bogoliubov-de Gennes equations.
To understand the physical meaning of the transformation (17), we 
express $\gamma_{n\uparrow}$ and $\gamma_{n\downarrow}$ by
the creation and destruction operators for an electron in the scattered 
state,$^{18}$  
\begin{eqnarray}
\gamma_{n\uparrow} &= &\sum_{n'}\bigl( u_{n,n'}^{*}c_{n'\uparrow} + v_{n,n'}c_{n'\downarrow}^{\dagger}\bigr),\nonumber \\
\gamma_{n\downarrow} &= &\sum_{n'}\bigl( u_{n,n'}^{*}c_{n'\downarrow} - v_{n,n'}c_{n'\uparrow}^{\dagger}\bigr),
\end{eqnarray}
where
\begin{eqnarray}
u_{n,n'} & =& \int \psi^{*}_{n'}({\bf r}) u_{n}({\bf r})d{\bf r}\nonumber \\
v_{n,n'} & =& \int \psi^{*}_{n'}({\bf r})v^{*}_{n}({\bf r})d{\bf r}. 
\end{eqnarray}

Accordingly, we obtain a vacuum state where $u_{n}({\bf r})\uparrow$ and 
$v_{n}^{*}({\bf r})\downarrow$ (instead of $\psi_{n}({\bf r})\uparrow$ and $
\psi_{\bar n}({\bf r})\downarrow$) are paired. 
They are the superpositions of the scattered states.
It is clear that the Bogoliubov-de Gennes equations cannot give rise to 
the correct Anderson's pairing because of the position dependence
of the pair potential. 
We need to supplement a pairing condition. 
If we assume a constant pair potential,$^{16,17}$ Anderson's
pairing is obtained. However, then, the impurity effect on the
phonon-mediated interaction is gone.

In the presence of magnetic impurities, even the constant
pair potential gives a pairing between the states which are the
linear superpositions of the scattered states.

\subsection{Homogeneous System }

As in Sec. IV. A, the unitary transformation generates a vacuum state 
and the self-consistency equation for the pairing between 
$u_{n}({\bf r})\uparrow$ and $v^{*}_{n}({\bf r})\downarrow$, (instead of
$\phi_{{\vec k}}({\bf r})\uparrow$ and  $\phi_{-{\vec k}}({\bf r})\downarrow$).   
Note that $u_{n}({\bf r})\uparrow$ and $v^{*}_{n}({\bf r})\downarrow$ are 
the linear superpositions of the plane wave states until
we constrain them. In this case, setting the pair potential 
gives a pairing between the plane wave states.
However, the kernel of the self-consistency 
equation has not been set accordingly.

\section{Pair Potential and Gap Parameter}
For a homogeneous system, it was shown$^{19}$
\begin{eqnarray}
\Delta({\bf r}-{\bf r'})=\int d{\vec k} e^{i{\vec k}\cdot({\bf r}-{\bf r'})}\Delta_{{\vec k}}.
\end{eqnarray}
But this relation is not exact because of the retardation
cutoff. 

Correct relation may be obtained only after incorporating the pairing constraint
into the self-consistency equation.
It is given
\begin{eqnarray}
\Delta({\bf r}-{\bf r'})=V\sum_{{\vec k}}{\Delta_{{\vec k}}\over 2E_{{\vec k}}}tanh
{E_{{\vec k}}\over 2T}\phi_{{\vec k}}({\bf r})\phi_{-{\vec k}}({\bf r'}).
\end{eqnarray}
Comparing Eq. (21) with the BCS gap equation and 
using Eq. (7), we also find  
\begin{eqnarray}
\Delta_{{\vec k}}=\int \phi_{{\vec k}}^{*}({\bf r})\phi^{*}_{-{\vec k}}({\bf r})
\Delta({\bf r})d{\bf r}.
\end{eqnarray}

In the presence of impurities, one finds that
\begin{eqnarray}
\Delta({\bf r})=V\sum_{n}{\Delta_{n}\over 2E_{n}}tanh
{E_{n}\over 2T}\psi_{n}({\bf r})\psi_{\bar n}({\bf r}),
\end{eqnarray}
and
\begin{eqnarray}
\Delta_{n}=\int \psi_{n}^{*}({\bf r})\psi^{*}_{\bar n}({\bf r})
\Delta({\bf r})d{\bf r}.
\end{eqnarray}
Eq. (24) was obtained first by Ma and Lee.$^{11}$

\section{Reinvestigation of Inhomogeneous Superconductors}
Now we need to reinvestigate the inhomogeneous superconductors
studied by Gor'kov's formalism or the Bogoliubov-de Gennes
equations.
In particular, Gor'kov's microscopic derivation$^{12}$ of the
Gizburg-Landau equation is not valid. The gradient term 
may not be derived microscopically.
To tackle the inhomogeneous problems, we must choose a correct
pairing and calculate the correct kernel for each
problem. Then the BCS theory may be more easy to apply. 
It is not clear whether the pair potential is the more
appropriate quantity than the gap parameter in inhomogeneous
superconductors. 

The following problems are required to restudy:

1. Dirty superconductors and localization

2. Magnetic impurities

3. Proximity effect, Andreev reflection and Josephson effect

4. Mesoscopic superconductivity 

5. Type II superconductors, vortex problem

6. Non-equilibrium superconductivity

7. High $T_{c}$ superconductors, heavy fermion superconductors

\subsection{Theory of Dirty Superconductors}

Table I lists the theories of impure superconductors.
Notice that Suhl and Matthias$^{20}$ and Abrikosov-Gor'kov$^{3}$
theories for $\Delta T_{c}$ are essentially equivalent.
Both theories 
over-estimate the change of the density of states caused by impurity scattering,
because they apply a retardation cutoff to 
the energy of the plane wave states and not of the scattered states.

\vskip 4pt
\centerline{{\bf TABLE I}. Theories of impure superconductors}
\vskip 4pt
\begin{tabular}{lll}\hline\hline
 & Ordinary impurity & Magnetic impurity\\ \hline
Anderson & $T_{c} =T_{co}$ \qquad & \\ 
AG & $T_{c} = T_{co}-{T_{co}\over \pi\omega_{D}\tau}({1\over \lambda}+{1\over 2})$  \qquad & $T_{c}=T_{co}-{\pi\over 4}{1\over \tau_{s}}$\\ 
Suhl and Matthias & $T_{c}\cong T_{co}-{T_{co}\over \lambda\omega_{D}\tau}$ \qquad & 
$T_{c}=T_{co}-{\pi\over 3.5}{1\over \tau_{s}}$\\ 
Baltensperger &  \qquad & $T_{c}=T_{co}-{\pi\over 4}{1\over \tau_{s}}$\\
Tsuneto & $T_{c}=T_{co}$ \qquad & \\
KO & $T_{c}=T_{co}-{T_{co}\over \pi\lambda E_{F}\tau}$ \qquad & $T_{c}
=T_{co}-{0.18\pi\over \lambda\tau_{s}}$\cr
& \qquad $-$ localization correction (Kim) \qquad & \\ \hline\hline
\end{tabular}
\vskip 8pt
Now we consider strong coupling theories of dirty
superconductors.
Tsuneto$^{6}$ obtained the gap equation
\begin{eqnarray}
\Sigma_{2}(\omega) = {i\over (2\pi)^{3}p_{o}}\int dq\int d\epsilon \int d\omega '
{qD(q,\omega-\omega')\eta(\omega')\Sigma_{2}(\omega')\over \epsilon^{2} - 
\eta^{2}(\omega')\omega'^{2}},
\end{eqnarray}
where $\eta=1 +{1\over 2\tau|\omega|}$, and $\tau$ is the collision time.
On the other hand, Kim$^{8}$ obtained a gap equation
\begin{eqnarray}
\Delta^{*}(\omega_{n}, m) = 
\sum_{n'}\lambda(\omega_{n}-\omega_{n'})
 \sum_{m'}V_{mm'}{\Delta^{*}(\omega_{n'},m')\over [-i\omega_{n'}
-\epsilon_{m'}][i\omega_{n'}-\epsilon_{m'}]},
\end{eqnarray}
where
\begin{eqnarray}
V_{mm'} = g^{2}\int |\psi_{m}({\bf r})|^{2}
 |\psi_{m'}({\bf r})|^{2}d{\bf r}.
\end{eqnarray}
Comparing Eqs. (25) and (26), we find that Tsuneto's result misses
the most important factor $V_{mm'}$, which gives the change of the
phonon-mediated interaction due to impurities. 
This factor is exponentially small for the localized
states. 

\subsection{Suppression of Magnetic Impurity Effect by Ordinary 
Impurities}

KO's results$^{7}$ can be explained physically.
People used to notice that the impurity effect is stronger
in the low $T_{c}$ material than in the high $T_{c}$ cuprate
materials. Because the size of Cooper pair is very much smaller in high $T_{c}$
material, it sees a very small number of the impurities and
so is not much influenced.  	
Consequently, $T_{c}$ change and its initial slope depend on
the material, which is predicted by KO.
However, the AG theory predicts the universal slope.
Because the AG theory pairs the states, which are the superpositions of the
normal states of the material, the very nature of the
material does not play an important role.

When the conduction electrons have a mean free path that is smaller than
the size of the Cooper pair (for a pure superconductor),
the effective size is reduced. Accordingly, if we add
ordinary impurities to the superconductors with
the magnetic impurities enough to reduce the size of the Cooper pair, 
the magnetic impurity effect is partially suppressed.
This compensation phenomena has been confirmed in several
experiments.$^{13-15}$

The notion of gapless superconductor is based on the misunderstanding
of the relation between the pair potential and the gap parameter.
>From Eq. (23), it is clear that the pair potential cannot be
finite when the gap parameter is zero.
It seems that the long range order between the magnetic
impurities or conduction electrons (especially in lead) caused
the gaplesslike behavior in the experiments. 
Then it would rather be called zero gap superconductors or
superconductors with nodes.

\subsection{Weak Localization Correction to the Phonon-mediated 
Interaction}
 
For the strongly localized states, the phonon-mediated interaction
is exponentially small like the conductance.
It is then expected that the same weak localization correction terms may occur 
in both quantities.
Using the wavefunction obtained by Mott and Kaveh,$^{21}$
it can be shown that$^{11}$ 
\begin{eqnarray}
V_{nn'}^{3d} &\cong& 
-V[1-{1\over (k_{F}\ell)^{2}}(1-{\ell\over L})],\nonumber \\
V_{nn'}^{2d} &\cong&  
 -V[1-{2\over \pi k_{F}\ell}ln(L/\ell)],\nonumber \\
V_{nn'}^{1d} &\cong&  
 -V[1-{1\over (\pi k_{F}a)^{2}}(L/\ell-1)],
\end{eqnarray}
where $a$ is the radius of the wire.

There are many experimental results which show the reduction of $T_{c}$
caused by weak localization.$^{22,23}$
Previously, it was interpreted by the enhanced Coulomb repulsion.
However, Dynes et al.$^{23}$ found a decrease of the Coulomb 
pseudo-potential $\mu^{*}$ with decreasing $T_{c}$.
We believe that this signals the importance of weak localization
correction to the phonon-mediated interaction.

\subsection{Other Inhomogeneous Superconductors}

We call attention to a few remarks against the conventional
real space formalism.

``By the use of the Gor'kov technique, Abrikosov and Gor'kov have succeeded
in $\cdots$, 
\quad it is {\sl entirely incorrect} as far as any physical results are concerned."$^{19}$

``Certain {\sl inconsistencies} are seen to develop with the use of
this approach (the de Gennes-Werthamer model), calling into
question previous results obtained."$^{24}$ 

``Is the discrepancy $\cdots$ indicative of the
certain {\sl fundamental inconsistencies} in the Green's function 
formulation of the theory of nonstationary superconductivity?"$^{25}$

\section{Conclusions}

It is shown that Gor'kov's formalism and  
the Bogoliubov-de Gennes equations need a pairing
constraint.
The resulting self-consistency equation is nothing but
another form of the BCS gap equation.
Most inhomogeneous superconductors should be reinvestigated.

I am grateful to Professor A. W. Overhauser for discussions. 
This work was supported by the National Science Foundation, Materials Theory
Program.

\end{document}